\documentclass[11pt]{revtex4}
\usepackage{amssymb,epsf}
\usepackage{latexsym}                              % 1 Sep 2013%

\begin{document}

\title{Asymptotically Lifshitz black hole solutions in F(R) gravity}
\author{S. H. Hendi$^{1,2}$\footnote{E-mail: hendi@shirazu.ac.ir},
B. Eslam Panah$^{3,4}$\footnote{E-mail: behzad\_eslampanah@yahoo.com} and C.
Corda$^{5}$\footnote{E-mail: cordac.galilei@gmail.com}}
\affiliation{$^1$ Physics Department and Biruni Observatory,
College of Sciences, Shiraz
University, Shiraz 71454, Iran \\
$^2$ Research Institute for Astrophysics and Astronomy of Maragha (RIAAM),
P.O. Box 55134-441, Maragha, Iran\\
$^3$ Department of Physics, University of Tabriz, Tabriz 51664, Iran\\
$^4$ Young Research Club, Islamic Azad University, Talesh Brach,
Talesh, Iran\\
$^5$ Institute for Theoretical Physics and Advanced Mathematics
Einstein-Galilei, Via Santa Gonda 14, 59100 Prato, Italy}

\begin{abstract}
We consider a class of spherically symmetric spacetime to obtain
some interesting solutions in $F(R)$ gravity without matter field
(pure gravity). We investigate the geometry of the solutions and
find that there is an essential singularity at the origin. In
addition, we show that there is an analogy between obtained
solutions with the black holes of Einstein-$\Lambda
$-power-Maxwell-invariant theory. Furthermore, we find that these
solutions are equivalent to the asymptotically Lifshitz black
holes. Also, we calculate $d^2F/dR^2$ to examine the
Dolgov-Kawasaki stability criterion.
\end{abstract}

\maketitle

\section{Introduction}

The nature of current accelerated expansion of the universe is one
of the mysterious and interesting topics in cosmology
\cite{Perlmutter}. In order to interpret this expansion, some
various candidates have been proposed by many authors, such as
cosmological constant idea \cite{Padmanabhan}, dark energy models
\cite{Copeland}, (exotic substance with large negative pressure
$P_{DE}\simeq -\rho _{DE}$ ($\rho _{DE}$ is the dark energy
density)) and modified gravities like Lovelock gravity
\cite{Lovelock}, brane world cosmology \cite{Gergely},
scalar-tensor theories \cite{Jordan} and also the so-called $F(R)$
gravity theories
\cite{FR,Cognola2008,Sotiriou,HendiPLB,HendiGRG,HendiPRD}.

On the other hand, amongst the nonlinear modifications of Einstein
gravity, $F(R)$ models seem to provide an interpretation to dark
energy, the hierarchy problem \cite{Cognola2006}, the four
cosmological phases \cite{Nojiri2006}, the power law early-time
inflation \cite{Starobinsky1980,Bamba2008}, late-time cosmic
accelerated expansion
\cite{Bamba2008,Carroll2004,Dombriz2006,Fay2007}, singularity
problem arising in the strong gravity regime
\cite{Abdalla,Briscese,Noj046006,Bamba,Kobayashi2008}, rotation
curves of spiral galaxies \cite{Capozziello2006} and detection of
gravitational waves \cite{Gwave}. In addition, when one considers
$F(R)$ theory as a modification of general relativity, it is quite
natural to ask about black hole existence in this theory.
Furthermore, it is notable that viable modifications in gravity
should pass all sorts of tests from the large scale structure of
the galaxy and cluster dynamics to the solar system tests. Hence,
many different models of $F(R)$ theory with various motivations
have been proposed (see \cite{HendiGRG} and references therein for
more details).

Moreover, regarding Einstein gravity, one can find a large number
of exact spherically symmetric solutions. One of important works
in the modified theories of gravity is obtaining the static
spherically symmetric solutions, a requirement usually referred to
as spatial isotropy and time independence. Therefore, the most
widely explored exact solutions in $F(R)$ gravity are the
spherically symmetric solutions. Spherically symmetric solutions
of $F(R)$ gravity have been studied before \cite{Multamaki2006}.
Furthermore, some of interesting black objects with various
geometry have been investigated in $F(R)$ gravity, for e.g.,
static and rotating black hole/string solutions, magnetic string
and so on \cite {Mazharimousavi2012,Larranaga}. Also, the black
hole/brane solutions with a nonlinear Maxwell source in $F(R)$
gravity has been investigated in \cite{Sheykhi2012}. Charged
spherically symmetric black hole solutions have been obtained in
Refs. \cite{HendiPLB,Mazharimousavi} and creation of the electric
charge and cosmological constant from pure gravity,
simultaneously, have been extracted in \cite{HendiGRG}.

Recently, Sebestiani and Zerbini \cite{Sebastiani} have considered
a class of interesting static spherically symmetric spacetime in
the form $ds^{2}=-\left(
\frac{r}{r_{0}}\right)^{q}g(r)dt^{2}+\frac{dr^{2}}{g(r)}+r^{2}d\Omega
_{k}^{2}$ with a model of $F(R)$ gravity. In this work, we would
like to consider this metric and obtain the exact spherically
symmetric black hole solutions for general value of $q$ and
discuss about the behavior of these solutions. We will consider
$F(R)=R+f(R)$ theory, namely pure gravity ($T_{\alpha \beta }=0$),
with a constant curvature scalar. One of our purposes is finding
the four-dimensional charged black hole solutions with
cosmological constant in pure gravity similar to solutions of
Einstein-$\Lambda$-power Maxwell invariant gravity
\cite{HendiEPJC2}. In addition, we will investigate the behavior
of these solutions and also discuss about Dolgov-Kawasaki (DK)
stability \cite{Dolgov}.

The outline of our paper is as follows. In section
\ref{FieldF(R)}, we obtain a static spherically symmetric
solutions of a class of $F(R)$ model and some special values of
free parameter, $q$, will be discussed in appendix. In section
\ref{Lifshitz}, we compare obtained solutions with asymptotically
Lifshitz black holes. We terminate our paper by a conclusion.

\section{ Exact black hole solutions \label{FieldF(R)}}

Considering the field equation of $4$-dimensional pure $F(R)$ gravity with
the following form \cite{FR,Cognola2008,Sotiriou,HendiPLB,HendiGRG,HendiPRD}
\begin{equation}
R_{\mu \nu }F_{R}-\nabla _{\mu }\nabla _{\nu }F_{R}+\left( \Box F_{R}-\frac{1%
}{2}F(R)\right) g_{\mu \nu }=0,  \label{FE}
\end{equation}
where ${R}_{\mu \nu }$ is the Ricci tensor, $F_{R}\equiv dF(R)/dR$
and $F(R)$ is an arbitrary function of Ricci scalar $R$. Here, we
want to obtain the static solutions of Eq. (\ref{FE}) with
positive, negative and zero curvature horizons. For this purpose,
we assume that the metric has the following form \cite{Sebastiani}
\begin{equation}
ds^{2}=-\left( \frac{r}{r_{0}}\right) ^{q}g(r)dt^{2}+\frac{dr^{2}}{g(r)}%
+r^{2}d\Omega _{k}^{2}  \label{Met1}
\end{equation}
where $q$ is an arbitrary constant and
\begin{equation}
d\Omega _{k}^{2}=\left\{
\begin{array}{cc}
d\theta ^{2}+\sin ^{2}\theta d\phi ^{2} & k=1 \\
d\theta ^{2}+\sinh ^{2}\theta d\phi ^{2} & k=-1 \\
d\theta ^{2}+d\phi ^{2} & k=0%
\end{array}
\right. .  \label{dOmega}
\end{equation}

To find the function $g(r)$, We use the components of Eq. (\ref{FE}) with
well-known $F(R)=R-\lambda \exp (-\xi R)+\kappa R^{n}$ model. This special
model and some of its properties have been investigated in Refs. \cite%
{HendiGRG,HendiPRD}. Using Eq. (\ref{FE}) with metric (\ref{Met1}), one can
obtain a general solution in the following form
\begin{equation}
g(r)=\frac{4k}{q^{2}+2q+4}+A r^{2}+Cr^{\Gamma _{1}+\Gamma _{2}}+Dr^{\Gamma
_{1}-\Gamma _{2}},  \label{gI}
\end{equation}
where $A$, $C$ and $D$ are integration constants and
\begin{eqnarray}
\Gamma _{1} &=&\frac{-3q}{4}-\frac{3}{2},  \label{Gamma1} \\
\Gamma _{2} &=&\frac{\sqrt{q^{2}+20q+4}}{4},  \label{Gamma2}
\end{eqnarray}

Since we know that the cosmological constant may arises from various models
of pure $F(R)$ gravity (see \cite{HendiGRG} and references therein), one may
set $A=-\Lambda/3$ to obtain a consistent solutions for $q=C=D=0$. In order
to satisfy all components of the field equation (\ref{FE}), we should set
the parameters of the $F(R)$ model to satisfy the following equations
\begin{equation}
\Lambda \Psi_{1} \left( \xi \lambda +e^{\xi R}\right) +6n \kappa R^{n}e^{\xi
R}=0,  \label{sol1}
\end{equation}
\begin{equation}
e^{\xi R}\left( \kappa R^{n} \Psi_{2}-4\Lambda \Psi_{3} \right) +\lambda
\left( \xi \Lambda \Psi_{4}+6\Psi_{1} \right) =0,  \label{sol2}
\end{equation}
where
\begin{eqnarray*}
\Psi_{1} &=&q^{2}+8q+24, \\
\Psi_{2} &=&6q^{2}(n-1)+24(n-2)(q+3), \\
\Psi_{3} &=&q^{3}+11q^{2}+48q+72, \\
\Psi_{4} &=&q^{4}+12q^{3}+68q^{2}+192q+288.
\end{eqnarray*}

Solving Eqs. (\ref{sol1}) and (\ref{sol2}) for $\lambda$ and
$\kappa$ with arbitrary $\xi $ lead to the following solutions
\begin{equation}
\lambda =\frac{R(n-1)e^{\xi R}}{n+\xi R},  \label{lambda1}
\end{equation}
\begin{equation}
\kappa =-\frac{\left( 1+\xi R\right) R^{1-n}}{n+\xi R}.
\label{kappa1}
\end{equation}

After some cumbersome manipulation we obtain
\begin{eqnarray}
&&R=\frac{\Lambda \Psi_{1} }{6}, \label{Ric1}\\
&&\lim_{r\longrightarrow 0}R_{\alpha \beta \gamma \delta
}R^{\alpha \beta
\gamma \delta }\longrightarrow \infty , \label{Riem1} \\
&&\lim_{r\longrightarrow \infty }R_{\alpha \beta \gamma \delta
}R^{\alpha \beta \gamma \delta }=\left[
\frac{q^{4}}{4}+2q^{3}+8q^{2}+16q+24\right] \frac{\Lambda ^{2}}{9}
\label{Riem2},
\end{eqnarray}
which confirm that there is a singularity at $r=0$. The position
of the black hole event horizon is determined as a largest root of
$g(r)=0$. Eqs. (\ref{Ric1}) and (\ref{Riem2}) show that the
curvature scalars are asymptotically finite and related to the
cosmological constant. But we cannot redefine a unique effective
cosmological constant for whole curvature scalars with an
arbitrary $q$. Therefore, we could not state these solutions are
asymptotically (a)ds with an effective cosmological constant.
Nevertheless, the asymptotic behavior of the solutions are near to
(a)ds solutions.

Here, we give a brief discussion about the stability of the
mentioned $F(R)$ model. It was showed that there is no stable
ground state for $F(R)$ models if $F(R)\neq 0$ and $F_{R}=0$
\cite{Schmidt}. For the obtained solutions, we find that
$F(R)=F_{R}=0$. Furthermore, it has been shown that
$F_{RR}=d^{2}F/dR^{2}$ is related to the the effective mass of the
dynamical field of the Ricci scalar \cite{Dolgov}. Therefore, the
positive effective mass, a requirement usually referred to as the
DK stability criterion, leads to the stable dynamical field
\cite{Faraoni}. In addition, One can discuss other instability
criteria of the asymptotically Lifshitz spacetimes considered in
\cite{Copsey,Stab2}. In Ref. \cite{Copsey}, the authors have
employed the initial value problem to investigate the stability of
such spacetimes. We will investigate other instability criteria
for the future works and in this paper, we consider DK stability.
In order to check the DK stability criterion, we should calculate
the second derivative of the $F(R)$ function with respect to the
Ricci scalar for this specific model
\begin{equation}
F_{RR}=-\frac{R(n-1)\left( \xi ^{2}+nR^{-2}(1+\xi R)\right) }{n+\xi R},
\label{FRR}
\end{equation}

\begin{figure}[tbp]
\epsfxsize=7cm \centerline{\epsffile{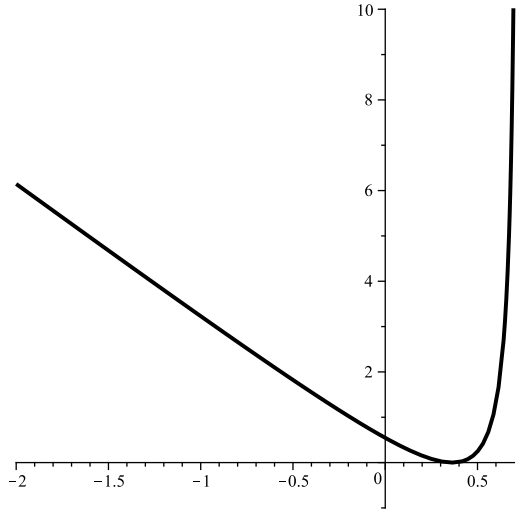}}
\caption{ $FRR$ versus $\protect\xi $ for $q=1$, $n=4$ and $\Lambda =-1$.}
\label{Fig1}
\end{figure}

The positivity of $F_{RR}$ depends on the model parameters. In order to
study the stability, we plot $F_{RR}$ in Fig. \ref{Fig1}. This figure shows
that one may obtain stable solution for special values of $\xi $ for
negative cosmological constant. In other words, we can set free parameters
to obtain positive $F_{RR}$.
It is notable to mention that since Eq. (\ref{FE}) is a trivial
relation for obtained solutions, we are not allowed to use the
conformal transformation to discuss about the dynamic of the
solutions.

\section{Asymptotically Lifshitz solutions:\label{Lifshitz}}

In recent years, it has been discussed about duality of gravity and
nonrelativistic scale invariant theories. It may be considered a scale
invariant fixed point that do not exhibit Galilean symmetry with the metric
of the corresponding gravity duals \cite{Kachru}. These metrics exhibit an
anisotropic scale-invariant field theories in which time and space scale
differently
\begin{eqnarray*}
t &\rightarrow &\lambda^{z}t, \\
\mathbf{x} &\rightarrow &\lambda \mathbf{x},
\end{eqnarray*}
which is characterized by the so-called dynamical exponent $z$. One may
obtain the standard scaling behavior of conformal invariant systems for $z=1$%
. In other words, for $z = 1$, it corresponds to relativistic invariance
(the scaling is isotropic). Also for arbitrary values of $z$, one can say
that the system has Lifshitz scaling. In some literature a special solution
was found, which corresponds to the asymptotically Lifshitz black hole with
various dynamical exponents \cite{AsympLif}.

Now, let us compare the presented solutions with asymptotically Lifshitz
black holes. In order to accomplish this goal, we consider the following
asymptotically Lifshitz ansatz
\begin{equation}
ds^{2}=-\left( \frac{r^{2}}{l^{2}}\right) ^{z}h(r)dt^{2}+\frac{l^{2}dr^{2}}{%
r^{2}h(r)}+r^{2}d\Omega _{k}^{2}.  \label{ALifMet}
\end{equation}
It is easy to show that Eq. (\ref{Met1}) is asymptotically Lifshitz
spacetime provided one apply the following transformation%
\begin{eqnarray}
&& r_{0} \longrightarrow l,  \nonumber \\
&& g(r) \longrightarrow \frac{r^{2}}{l^{2}}h(r),  \nonumber \\
&& q \longrightarrow 2z-2.  \label{Trans}
\end{eqnarray}
In other word, asymptotically Lifshitz metric (\ref{ALifMet})
(with the mentioned $F(R)$ model) leads to the following metric
function
\begin{equation}
h(r)=\frac{kl^{2}}{\left( z^{2}-z+1\right) r^{2}}-\frac{\Lambda l^{2}}{3}%
+Cl^{2}r^{\Gamma _{1}+\Gamma _{2}-2}+Dl^{2}r^{\Gamma _{1}-\Gamma _{2}-2},
\label{h(r)}
\end{equation}
where
\begin{eqnarray}
\Gamma _{1} &=&-\frac{3z}{2},  \label{G1} \\
\Gamma _{2} &=&\frac{\sqrt{z^{2}+8z-8}}{2}.  \label{G2}
\end{eqnarray}

It is notable that the generic solution Eqs. (\ref{ALifMet}) and (\ref{h(r)}%
) reduces to the solution of obtained in Ref. \cite{MannL}, for $C=D=0$ and $%
z=2$.

\section{Conclusions}

In this paper, we have considered a new class of static
spherically symmetric spacetime with a special model of $F(R)$
gravity. We have obtained some interesting solutions for different
values of free parameter, $q$, and studied the geometrical
properties of the solutions.

We have investigated the analogy between obtained solutions with $4$%
-dimensional solutions of Einstein-$\Lambda $ gravity in the presence of a
nonlinear source, namely power Maxwell invariant (where $s$ denotes its
nonlinearity parameter). In other words, one can find that the integral
constant $C$ may interpreted as $(charge)^{2s}$ provided $s=1/2-(\Gamma
_{1}+\Gamma _{2})^{-1}$. In addition, for $s=1/2-(\Gamma _{1}-\Gamma
_{2})^{-1}$, one may interpreted $D$ as $(charge)^{2s}$.

Furthermore, we have calculated Kretschmann scalar and found that
there is a curvature singularity at $r=0$. We have investigated
the behavior of the solutions for large value of $r$
($r\longrightarrow \infty $) and showed that the asymptotic
behavior of the solutions is neither flat nor (a)dS. Also, we have
examined the DK stability criterion and showed that one may obtain
a stable theory provided the parameters of the $F(R)$ model are
chosen suitably.

Finally, we should remark that obtained solutions are interesting
because of the following three main reasons: (i) both the geometry
of the horizon(s) and asymptotic behavior of the solutions depend
on the values of free parameter $q$, (ii) starting from pure
$F(R)$ gravity and extract the solutions of Einstein-$\Lambda
$-power Maxwell invariant theory, (iii) obtained solutions are,
exactly, the same as asymptotically Lifshitz black holes, provided
the transformation between them are chosen suitably.

\section{Appendix}

Now, we would like to discuss about the geometry of the spacetime with
various values of metric parameters.

\subsection{First type: $C=D=0$}

Considering the first two terms of Eq. (\ref{gI}), one can obtain
interesting spacetimes with some special values of $q$.

\subsubsection{Case I: $q=0$}

In this trivial case, the Ricci and Kretschmann scalars are $4\Lambda $ and $%
\frac{8\Lambda ^{2}}{3}$, respectively, and therefore the solution is
asymptotically (a)dS.

\subsubsection{Case II: $q=-2$}

Inserting $q=-2$ in Eq. (\ref{gI}), one can obtain
\begin{equation}
ds^{2}=-\left( \frac{r_{0}}{r}\right) ^{2}\left( k-\frac{\Lambda }{3}%
r^{2}\right) dt^{2}+\frac{dr^{2}}{\left( k-\frac{\Lambda }{3}r^{2}\right) }%
+r^{2}d\Omega _{k}^{2},  \label{Met2}
\end{equation}
with the following Ricci and Kretschmann scalars
\begin{eqnarray}
R &=&2\Lambda ,  \label{Ricci(q=-2)} \\
R_{\alpha \beta \gamma \delta }R^{\alpha \beta \gamma \delta } &=&\frac{%
72k^{2}}{3r^{4}}+\frac{4\Lambda ^{2}}{3},  \label{Kresch(q=-2)}
\end{eqnarray}
where confirm that the spacetime is asymptotically near to (a)dS which its
curvature scalars are the half of their counterpart in (a)dS spacetime.

\subsubsection{Case II: $q=-4$}

One may consider $q=-4$ in Eq. (\ref{gI}) to obtain%
\begin{equation}
ds^{2}=-\left( \frac{r_{0}}{r}\right) ^{4}\left( \frac{k-\Lambda r^{2}}{3}%
\right) dt^{2}+\frac{3dr^{2}}{k-\Lambda r^{2}}+r^{2}d\Omega _{k}^{2},
\label{Met3}
\end{equation}
where the Ricci and Kretschmann scalars are
\begin{eqnarray}
R &=&\frac{4\Lambda }{3},  \label{Ricci(q=-4)} \\
R_{\alpha \beta \gamma \delta }R^{\alpha \beta \gamma \delta } &=&\frac{%
8\Lambda ^{2}}{3}-\frac{64\Lambda k}{9r^{2}}+\frac{64k^{2}}{3r^{4}}
\label{Kresch(q=-4)}
\end{eqnarray}
This solution is near to asymptotically (a)dS spacetime, but as
one may find, obtained scalars are different delicately. Now, we
investigate the solutions with nonzero $C$ and $D$.

\subsection{Second type: $C\neq 0$ \& $D\neq 0$}

In order to have a real solution with nonzero $C$ or $D$, we
should consider $q\leqslant -10-4\sqrt{6}\simeq -19.8$ or
$q\geqslant -10+4\sqrt{6}\simeq -0.2$ (in other words: $q\notin
\left( -10-4\sqrt{6},-10+4\sqrt{6}\right)$). In general,
considering the metric (2) with $g(r)=A r^m$, one can show that
the curvature scalars diverge at $r=0$ with arbitrary $m$. In
addition, curvature scalars diverge at spatial infinity for $m>2$.
To avoid non-physical singularity, we restrict our discussions for
$m \leq 2$ and hence we only consider $q \geq -10+4\sqrt{6}$
branch ( for a discussion about the essential singularity at
infinity, we refer the reader to Ref. \cite{Elsner}).

Now, we discuss about some specific values of allowed $q$. At
first, we consider the trivial case ($q=0$) and then we
investigate the boundary values of $q$ and after that we analyze
some interesting values of it which lead to integer number for the
numerator of Eq. (\ref{Gamma2}).

\subsubsection{Case I: $q=0$}

This trivial case leads to

\begin{eqnarray*}
\Gamma _{1} &=&-3/2, \\
\Gamma _{2} &=&1/2,
\end{eqnarray*}
with trivial RN solution as follows
\begin{equation}
ds^{2}=-g(r)dt^{2}+\frac{dr^{2}}{g(r)}+r^{2}d\Omega _{k}^{2},
\label{metCaseI}
\end{equation}
where
\begin{equation}
g(r)=k-\frac{\Lambda r^{2}}{3}-\frac{M}{r}+\frac{Q^{2}}{r^{2}},
\label{gCaseI}
\end{equation}
where we redefined $C=-M$ and $D=Q^{2}$. This solution indicates that one
may obtain cosmological constant as well as electric charge of Maxwell field
from pure gravity which has studied before \cite{HendiGRG}.

\subsubsection{Case II: $q=-10+4\protect\sqrt{6}$}

\begin{figure}[tbp]
\epsfxsize=7cm \centerline{\epsffile{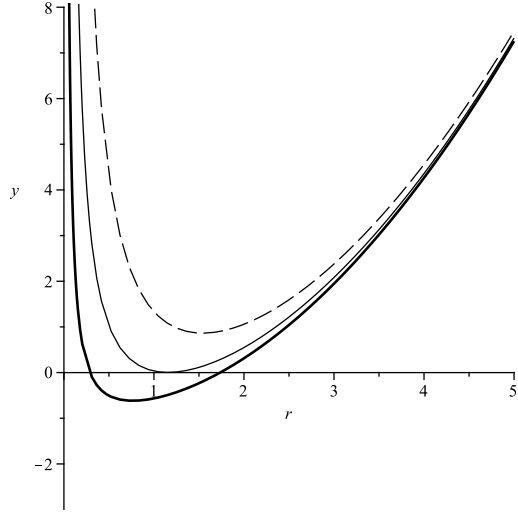}}
\caption{ $g(r)$ (Eq. (\protect\ref{g3})) versus $r$ for $k=-1$, $D=0.1$, $%
\Lambda =-1$ and $C=0.1$ (bold line), $C=0.7$ (solid line) and $C=2$ (dashed
line).}
\label{Fig3}
\end{figure}

For this value of $q$, we encounter with vanishing $\Gamma _{2}$ and $\Gamma
_{1}=\left( 6-3\sqrt{6}\right) \simeq -1.35$, and so the solution is%
\begin{equation}
g(r)\approx \frac{4k}{3.64}-\frac{\Lambda r^{2}}{3}+\frac{(C+D)}{r^{1.35}},
\label{g3}
\end{equation}
where the second term is dominant for large $r$. Calculating the Ricci and
Kretschmann scalars show that they diverge at the origin and they are finite
for $r\neq 0$. It is easy to show that the presented solutions may be
interpreted as black hole solutions with two horizons, extreme black hole
and naked singularity provided the parameters of the solutions are chosen
suitably (see Fig. \ref{Fig3} for more details). Considering $k=0$, it is
easy to show that the mentioned $g(r)$ is the same as that in the
Einstein-power Maxwell invariant gravity \cite{HendiEPJC2} with vanishing
mass when the nonlinearity parameter is chosen $s=\frac{3\sqrt{6}-4}{6\left(
\sqrt{6}-2\right) }$.

\subsubsection{Case III: $q=1$}

One may consider another interesting value of $q$ to obtain%
\begin{eqnarray*}
\Gamma _{1} &=&\frac{-9}{4}, \\
\Gamma _{2} &=&\frac{5}{4}.
\end{eqnarray*}%
where Eq. (\ref{gI}) simplified to%
\begin{equation}
g(r)=\frac{4k}{7}-\frac{\Lambda r^{2}}{3}+\frac{C}{r}+\frac{D}{r^{7/2}}.
\label{gII}
\end{equation}%
Despite of the first constant term in Eq. (\ref{gII}), this
solution is near to asymptotically (a)dS charged solution of
nonlinear Maxwell gravity \cite{HendiEPJC2} with $s=11/14$ (for
$k=0$, they are exactly the same). Thus, one may conclude that $C$
and $D$ are related to the mass and charge of the spacetime,
respectively. In other words, we can extract the electric charge
and cosmological constant form pure gravity, simultaneously. One
may calculate the Ricci and Kretschmann scalars to achieve
\begin{eqnarray}
R &=&\frac{11}{2}\Lambda ,  \label{Rq1} \\
\left. R_{\alpha \beta \gamma \delta }R^{\alpha \beta \gamma \delta
}\right\vert _{\text{Large }r} &=&\frac{67\Lambda ^{2}}{12}-\frac{2k\Lambda
}{7r^{2}}+O(r^{-4}),  \label{RR1q1} \\
\left. R_{\alpha \beta \gamma \delta }R^{\alpha \beta \gamma \delta
}\right\vert _{\text{Small }r} &=&\frac{141D^{2}}{r^{11}}+\frac{22CD}{%
r^{17/2}}-\frac{104kD}{7r^{15/2}}+\frac{6C^{2}}{r^{6}}+O(r^{-11/2}).
\label{RR2q1}
\end{eqnarray}%
Equations (\ref{Rq1})-(\ref{RR2q1}) show that there is an
essential singularity at the origin and this solution is,
asymptotically, near to (a)dS spacetime.

\begin{figure}[tbp]
\epsfxsize=7cm \centerline{\epsffile{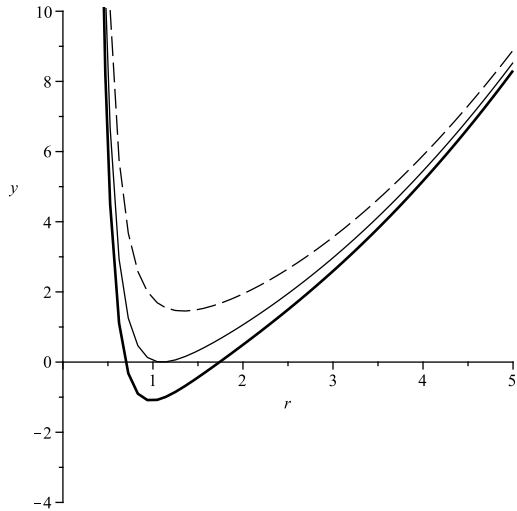}} \caption{
$g(r)$ (Eq. (\protect\ref{gII})) versus $r$ for $\Lambda =-1$,
$D=1 $, $k=1$, and $C=-0.1$ (dashed line), $C=-1.86$ (solid line),
and $C=-3$ (bold line).} \label{Fig4}
\end{figure}

In order to investigate the geometry of the function $g(r)$, we
plot it versus $r$ in Fig. \ref{Fig4}. This figure shows that the
geometry of the horizons are near to those in charged black holes.

\subsubsection{Case IV: $q=4$}

For $q=4$, and therefore $\Gamma _{1}=-9/2$ and $\Gamma
_{2}=10/4$, Eq. (\ref{gI}) reduces to
\begin{equation}
g(r)=\frac{k}{7}-\frac{\Lambda r^{2}}{3}+\frac{C}{r^{2}}+\frac{D}{r^{7}}.
\label{gIII}
\end{equation}
Let us first consider $D=0$. In this case Eq. (\ref{gIII}) may be
interpreted as approximately asymptotically (a)dS charged solution with
vanishing mass and\ therefore $C$ is related to electric charge.

In addition, one may consider $C=0$ with nonvanishing $D$ to obtain
approximately asymptotically (a)dS solution that for $k=0$, this solution is
the same one obtained in Ref. \cite{HendiEPJC2} when the nonlinearity
parameter is equal to $9/14$. We can obtain the Ricci and Kretschmann
scalars in the following manner
\begin{eqnarray}
R &=&12\Lambda , \\
R_{\alpha \beta \gamma \delta }R^{\alpha \beta \gamma \delta } &=&\frac{%
136\Lambda ^{2}}{3}-\frac{64k\Lambda }{7r^{2}}+\frac{\frac{32C\Lambda }{3}+%
\frac{192k^{2}}{49}}{r^{4}}-\frac{32kC}{7r^{6}}+\frac{24C^{2}}{r^{8}}-\frac{%
176\Lambda D}{r^{9}}+  \nonumber \\
&&\frac{48kD}{7r^{11}}-\frac{32CD}{r^{13}}+\frac{444D^{2}}{r^{18}}.
\end{eqnarray}
Therefore as we expected, there is a curvature singularity at
$r=0$, and asymptotic behavior of the spacetime is near to (a)dS.

\begin{center}
acknowledgements
\end{center}
The authors thank an unknown referee for good comments and advices
which permitted to improve this paper. S. H. Hendi wishes to thank
Shiraz University Research Council. This work has been supported
financially by Research Institute for Astronomy \& Astrophysics of
Maragha (RIAAM), Iran.

\end{document}